# Coherent dynamics and ultrafast evolution of optical nonlinearity of self-hybridized exciton-polaritons in $WSe_2$

Niu Xu[1,2+], Jin Yang[1,2+], Kaizhen Liu[1,2], Jianghao Shen[1,2], Weiming Song[1,2], Abbas Chimeh[3,4], Daniel Timmer[5], Jin-Hui Zhong[1,2]*

[1]State Key Laboratory of Quantum Functional Materials, Shenzhen Key Laboratory of Phononics and Intelligent Thermal Materials, Department of Materials Science and Engineering, Southern University of Science and Technology, Shenzhen 518055, China

[2]Guangdong Provincial Key Laboratory of Sustainable Biomimetic Materials and Green Energy, Institute of Innovative Materials, Southern University of Science and Technology, Shenzhen 518055, China

[3]Quantum Matter Lab, Department of Physics, College of Science, University of Tehran, Tehran, Iran

[4]Quantum Machine Learning Lab, Department of Physics, College of Science, University of Tehran, Tehran, Iran

[5]Fachbereich Physik, Philipps-Universität Marburg, Marburg 35032, Germany

[+]These authors contribute equally to this work.

*Email: zhongjh@sustech.edu.cn

## Abstract

Exciton-polaritons (EPs) are valuable for investigating macroscopic quantum coherent phenomena. However, overcoming the rapid decoherence of EPs under ambient conditions is challenging, even using delicately fabricated optical cavities. Here, we show pronounced coherent dynamics in thin flakes of $WSe_2$ supporting self-hybridized EPs. Ultrafast two-dimensional electronic spectroscopy (2DES) resolves two types of quantum beating: Rabi oscillation with a period of ~ 24 fs and a slower, ~ 50 fs period oscillation, revealing coherent coupling not only between excitons and cavity, but also between polaritons and dark states. Rapid evolution of 2DES line shape indicates a change of dominant optical nonlinearities from Rabi contraction of polaritons to many-body interactions of excitons in the first 100 fs, marking the transition from coherent to incoherent polariton excitations. The results are rationalized by a three-coupled oscillator model considering the inhomogeneous cavity field distribution. We also demonstrate coherent quantum beating between A exciton and its excited Rydberg state A* under ambient conditions. Our results shed light on the coherent control in readily accessible semiconductor thin flakes.

## Introduction

Coherent phenomena in strongly coupled light-matter systems have attracted considerable research interest, holding potential in Bose-Einstein condensation,[1] quantum-fluid and superfluid phenomena,[2] low-threshold polariton lasing,[3] ultrafast all-optical switching and modulation,[4] and quantum simulation of many-body systems.[5] Macroscopic coherence, manifested by the macroscopic occupation of a coherent quantum state, underlies collective phenomena such as polariton condensation and superfluidity.[6, 7] Exciton-polaritons (EPs), arising from the strong coupling between optical modes and excitonic transitions, provide a versatile platform for realizing these phenomena, which are fundamentally rooted in quantum coherence. In EP systems, the hybridization of photonic and material excitations, together with exciton-mediated interactions, gives rise to coherent dynamics and pronounced optical nonlinearities.[8-11] These coherent dynamics provide a basis for the controlled manipulation of EPs.

Quite often, to achieve strong coupling under ambient conditions, the optical electric field must be sufficiently enhanced.[12, 13] Various cavity structures such as plasmonic nanostructures,[14] photonic crystals,[15] and Fabry-Pérot (FP) resonators[16] are commonly used, which generally demands delicate control of the cavity structural parameters, often entailing complicated fabrication procedures. Instead, optical eigenmodes exist in objects with a refractive index different from their surroundings. Even in the absence of an external cavity, these eigenmodes can potentially strongly couple with material excitons to form self-hybridized EPs.[17] Examples of such "cavities" include FP resonances in finite-thickness flakes,[18-21] bound or quasi-bound states in the continuum in periodic or metasurface structures,[22, 23] localized Mie or anapole resonances in nanostructures,[24-26] as well as guided modes in waveguide geometries.[27, 28] Some of these cavities, for instance FP modes supported in thin flakes, significantly eliminate the complexity in cavity fabrication. These inherently open platforms are suitable for electrical injection, electric/strain/magnetic-field tuning, and on-chip integration, offering promising routes toward tunable and scalable devices based on self-hybridized EPs.[29-31] The exposure of active materials to vacuum, air or solution, may also benefit their applications in, for example, polariton chemistry.[32]

However, the origin of optical nonlinearity and the ultrafast dynamics of self-hybridized EPs, particularly the coherent dynamics, remain unexplored.[10, 33, 34] Probing such coherent dynamics of EPs, especially in the electronic transition regime, is challenging because coherence is limited by dissipation and dephasing processes.[35] Furthermore, the mechanisms of polariton dephasing and the interaction between polariton and dark states are largely unknown. These insights would be vital for exploring the potential of self-hybridized EPs for coherent optoelectronic applications.

Ultrafast two-dimensional electronic spectroscopy (2DES) is an invaluable tool for probing coherent interactions and dynamics.[36] The unique spectral features provided by 2DES, including diagonal peaks, cross-peaks, spectral line shapes and temporal evolution, can provide a comprehensive picture of how energy flows and coherence is lost.[37-43] Recent studies of molecular J-aggregates and monolayer $WS_2$ strongly coupled to plasmonic resonators,[10, 44] together with self-hybridized EPs in transition metal dichalcogenide (TMD) metasurfaces,[9] have demonstrated the advantage of 2DES for investigating the coherent dynamics of EPs.

Herein, we use 2DES to dissect the coherent dynamics and evolution of optical nonlinearities of self-hybridized EPs in $WSe_2$ flakes. We resolved two types of quantum beating, Rabi oscillation with a period of ~ 24 fs and a slower, ~ 50 fs period oscillation, revealing coherent coupling not only between excitons and cavity, but also between polaritons and dark states. Concurrently, rapid evolution of 2DES line shape indicates a change of dominant optical nonlinearities from Rabi contraction of polaritons to many-body interactions of excitons in the first 100 fs, marking the transition from coherent to incoherent polariton excitations. We rationalize the result by a three coupled oscillator model with two types of excitons that couple strongly and weakly to the same cavity mode. We also demonstrate coherent quantum beating between A exciton and its excited Rydberg state A* under ambient conditions. Our results indicate that quantum coherence can be readily achieved in self-hybridized EPs without delicately fabricated cavity structures, significantly expanding the scope of candidate materials for coherent manipulation.

**Self-hybridized exciton-polaritons in a $WSe_2$ flake**

We prepared $WSe_2$ flakes with different thicknesses on a gold substrate (Fig. 1a and Supplementary Fig. 1). For a 60 nm thick flake (Fig. 1b), an incidence angle ( $\theta$ )-independent resonance is observed at 1.64 eV, attributed to the A exciton (AX) of $WSe_2$.[45] When the thickness is increased to 190 nm, the optical responses are dramatically different (Fig. 1c). We barely see a peak at the AX energy. Instead, two peaks at ~1.55 eV and ~1.72 eV are observed. Fig. 1d compares the reflection spectra of the two flakes at $\theta = 5.7°$. For a $WSe_2$ flake with sub-wavelength thickness and high refractive index ($n \sim 4.6$, in the visible wavelength range),[46] multiple reflections at the top and bottom interfaces can form FP cavity modes that couple to excitons in $WSe_2$.[20]

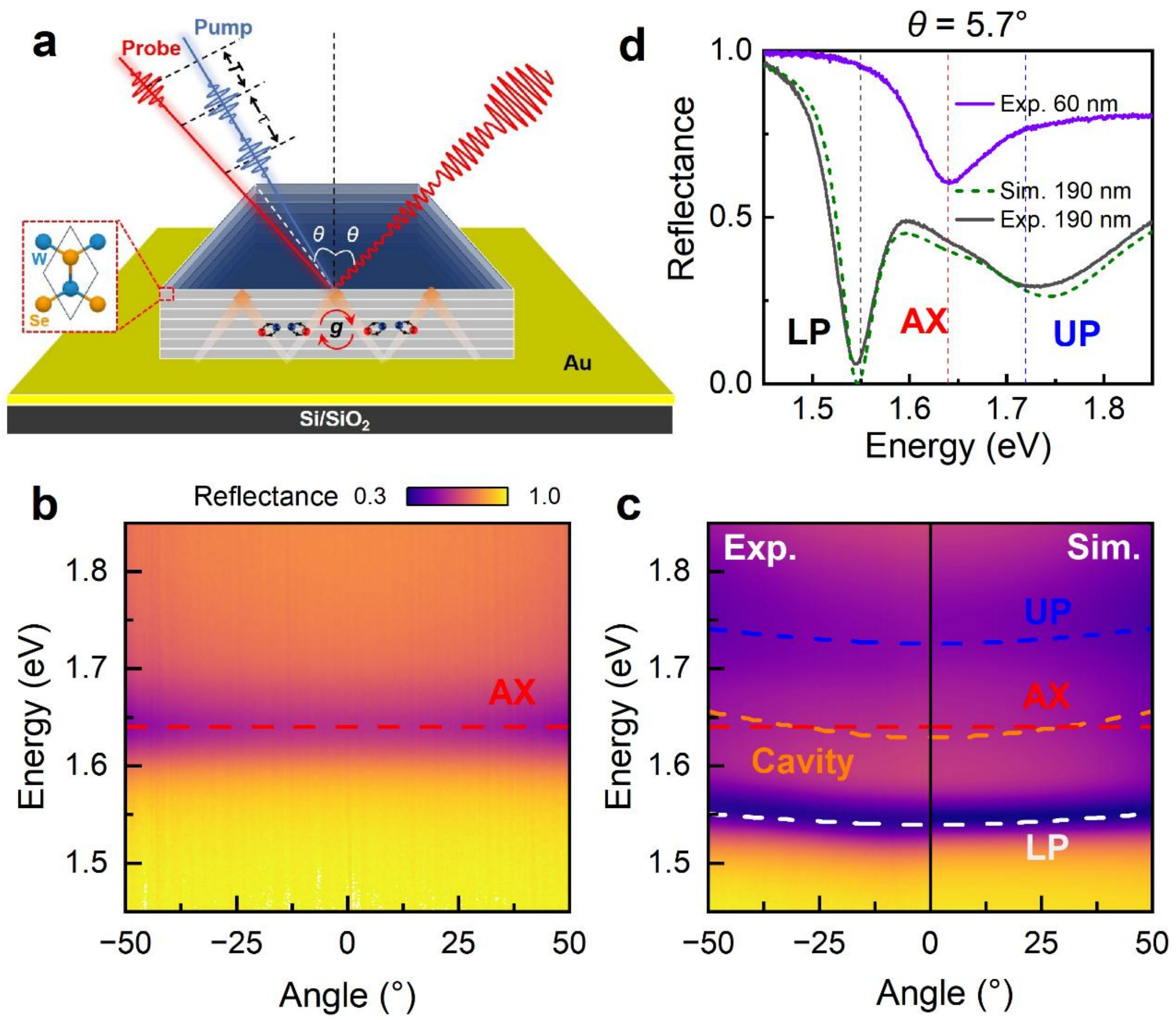

**Fig. 1: Self-hybridized exciton polaritons in $WSe_2$ flakes.**

**a** Scheme of the sample and 2DES experiment. The flake is pumped by a phase-locked pump pulses pair at an incidence angle of $\theta = 5.7°$, separated by coherence time τ and interrogated by a probe pulse with the same incidence angle after the waiting time $T$. For ultrafast transient reflection spectroscopy, τ = 0 fs. **b** Experimental angle-resolved reflectance spectra of the 60 nm thick $WSe_2$ flake. **c** Experimental (left) and TMM simulated (right) angle-resolved reflectance spectra of the 190 nm thick $WSe_2$ flake. The dashed lines are the fitting results by the COM. **d** Reflectance spectra measured at an incidence angle $\theta = 5.7°$ of the 190 nm thick (black) and the 60 nm thick (violet) $WSe_2$ flakes. Dashed olive line shows the simulated reflectance spectrum by TMM. A clear splitting of peaks is seen in the 190 nm thick flake.

A transfer matrix method (TMM) simulation was performed to understand the thickness dependent FP cavity modes assuming a dielectric layer with constant refractive index $n$ = 4.6 and no exciton resonance (Supplementary Fig. 2). This indicates that the 2nd FP cavity mode ($E_{Cavity}^{2nd} \approx 1.63\ eV$) is almost resonant with AX for the 190 nm thick flake. Then, by using the complex refractive index including the excitonic response[46], we theoretically obtain the angle-resolved reflectance spectra of a 190 nm thick flake (Fig. 1c right and olive dashed line in Fig. 1d) with TMM, reproducing the experiment well. The energy dispersion is further fitted by a coupled oscillator model (COM) and the results are plotted as dashed lines in Fig. 1c, revealing a coupling strength of $g \approx 90$ meV. This satisfies $g > \sqrt{\frac{1}{2}(\gamma_X^2 + \gamma_C^2)}$, with exciton dephasing rate ($\gamma_X$ = 35 meV) and cavity decay rate ($\gamma_C$ = 47 meV), see Supplementary Fig. 3 for details, indicating that the system is in the strong coupling regime[10, 47]. Therefore, the peaks at ~1.55 eV and ~1.72 eV can be assigned as the lower (LP) and upper polariton (UP) branches, respectively. The fractions of photons and excitons are both within 50% ± 2%, i.e., the polariton is roughly half-photon half-exciton for $\theta = 0 - 50°$ (Supplementary Fig. 4). Since there is no top mirror, the polaritons are formed by self-hybridization between excitons and the FP cavity mode.[21]

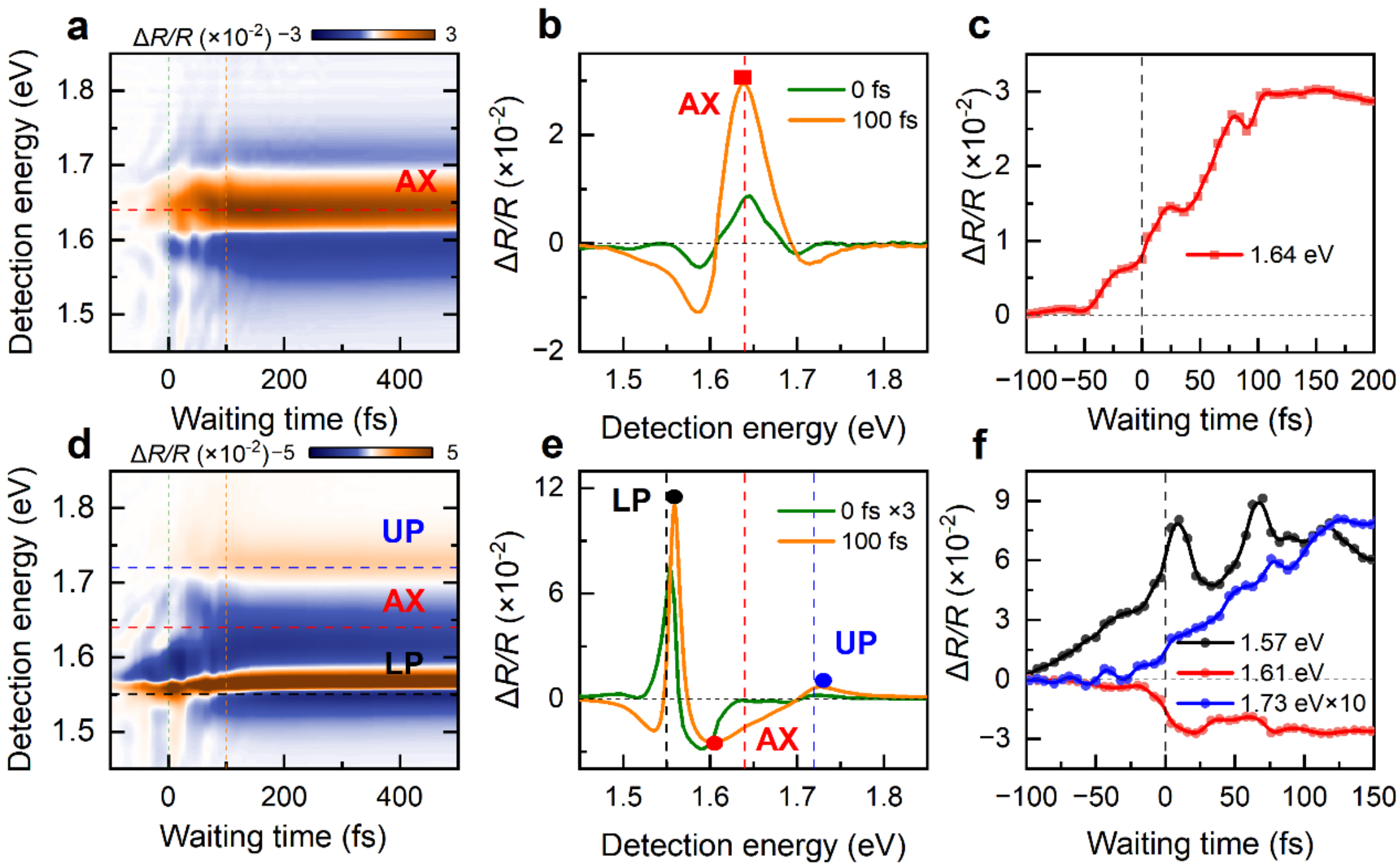

**Fig. 2: Ultrafast transient reflection (Δ*R/R*) spectra.**

**a** Two-dimensional plot of Δ*R/R* spectra of the 60 nm thick $WSe_2$ flake, showing a main positive peak near the AX resonance (red dashed line), accompanied with two negative side lobes on both low and high energy sides. **b** Δ*R/R* spectra at $T = 0$ fs and $T = 100$ fs from (**a**). The vertical dashed line marks the AX energy. **c** Time dynamics taken at the positive peak maximum marked by a red rectangle in (**b**). **d** Two-dimensional plot of Δ*R/R* spectra of the 190 nm thick $WSe_2$ flake. The Δ*R/R* spectra exhibit two main negative-positive peaks originating from LP and UP (black and blue dashed lines), and a strong negative ESA signal close to the bare AX peak position (red dashed line). **e** Δ*R/R* spectra at $T = 0$ fs and $T = 100$ fs from (**d**). The line shape undergoes drastic changes within 100 fs, indicating rapid variations in optical nonlinearity. **f** Time dynamics taken at the energies marked by the black (LP), blue (UP) and red spots in (**e**). Temporal modulations are seen.

We used transient reflection (TR) spectroscopy to study the dynamics of the 60 nm (Fig. 2a) and 190 nm thick (Fig. 2d) flakes. Representative spectra at $T = 0$ fs and 100 fs for the two samples are plotted in Fig. 2b and

Fig. 2e. For the 60 nm thick flake, we observe a positive peak centered at detection energy ($E_{de}$) of 1.64 eV, i.e., the AX energy. The spectra also show two negative side lobes on the high and low energy sides of AX. It has been well documented that the optical nonlinearity of TMD excitons is governed by many-body interactions (MBIs) of excitons, including excitation induced dephasing (EID), Pauli blocking (PB), and excitation induced shift (EIS), etc.[10, 33, 48, 49] Their line shapes are summarized in Supplementary Fig. 5. The negative-positive-negative line shape is characteristic of EID. The negative wing at low energy is stronger than that at high energy, indicating a finite EIS effect, specifically a red shift upon pump excitation.[49, 50] The dynamics of the AX peak show faint modulations (Fig. 2c), as further discussed later.

The spectra of the 190 nm thick flake are distinctly different (Fig. 2d, 2e). For $T$ = 0 fs, the line shape around LP is dispersive, i.e., positive-negative. The signal around UP is also dispersive but with a flipped sign. This is a typical feature of Rabi contraction, i.e., a reduction of Rabi splitting due to the decrease in number of participating excitons in polaritons after pump excitation.[16, 51-53] The energy of LP and UP blue and red shift, respectively, leading to dispersive line shapes. Fundamentally, this results from PB (Supplementary Fig. 5a). For $T$ = 100 fs, the line shape transforms into negative-positive-negative-positive features. Fig. 2f shows peak dynamics at three energies marked by circles with corresponding colors in Fig. 2e, exhibiting temporal modulations. The periods are, however, much lower than the expected Rabi oscillation of ~ 22 fs.

## Coherent coupling and dynamics probed by 2DES

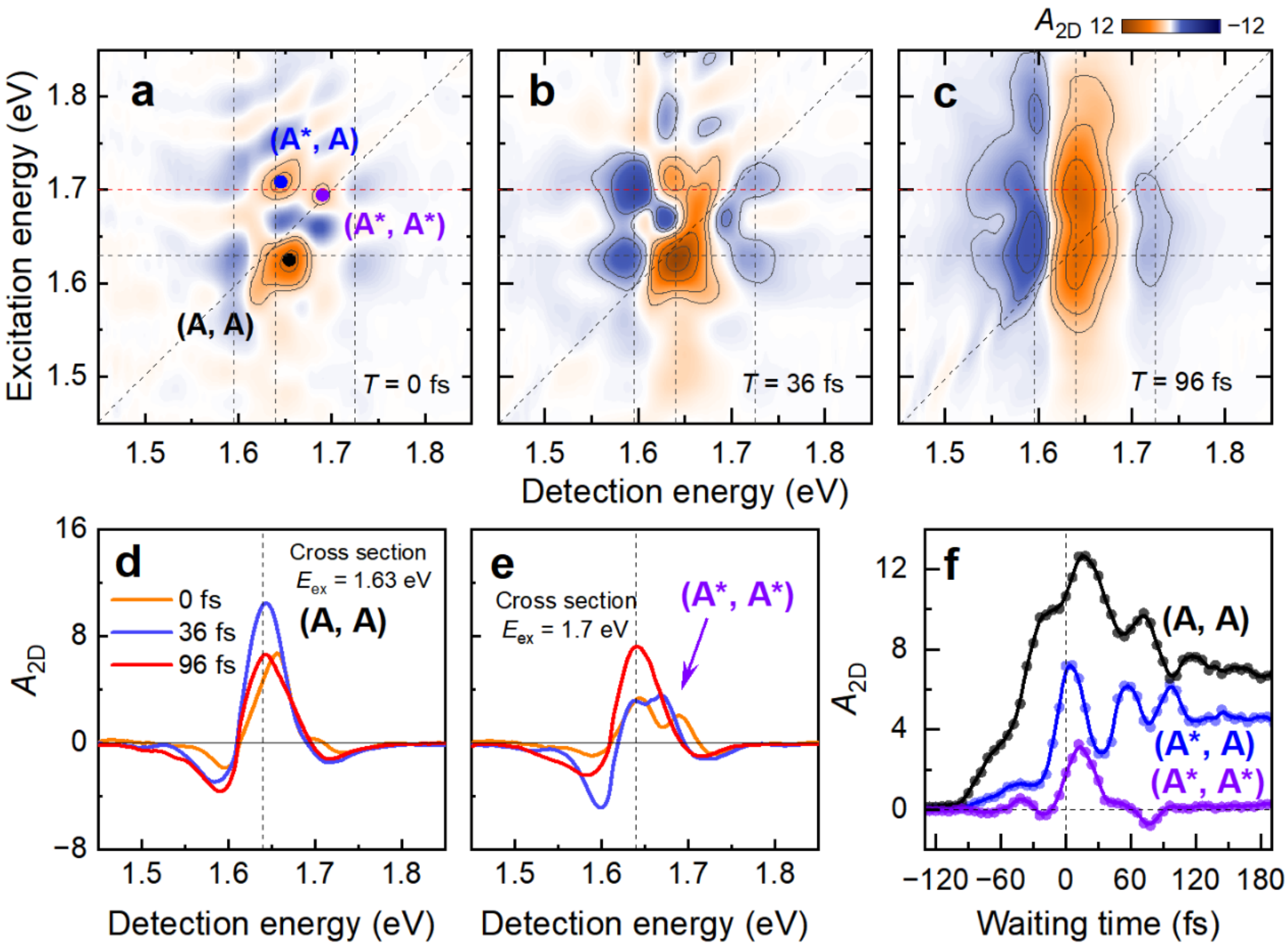

**Fig. 3 2DES of the 60 nm thick $WSe_2$ flake.**

**a-d** 2DES maps at selected waiting times of *T*= 0 fs (**a**), 36 fs (**b**) and 96 fs (**c**). **d-e** Cross section along the detection energy for excitation energy of 1.63 eV (**d**) and 1.7 eV (**e**). **f** Waiting time dynamics of (A, A), (A*, A*) diagonal peaks and (A*, A) cross peak, taken from the positions marked by the black, violet and blue circles in (**a**), respectively. All dynamics exhibit pronounced coherent oscillation with a period of ~ 50 fs (corresponding to ~80 meV).

To get deeper understanding of the nonlinearities and dynamics of excitons and EPs, we further performed 2DES measurements (Methods). Different from TR, a phase-locked pump pulse pair replaces the single pump pulse (Fig. 1a). By scanning the time interval $\tau$ (coherence time) between the two pump pulses and performing Fourier transform (FT), we get excitation energy, providing 2DES maps correlating excitation and detection energies. 2DES maps of the 60 nm thick flake at several waiting times are presented in Fig. 3a-c.

For $T = 0$ fs and at $E_{ex} = 1.63$ eV, i.e., almost resonant with AX, we see negative-positive-negative line shape peaks, evidenced by the cross section in Fig. 3d. The low energy negative wing is stronger than that at the high energy side, indicating a combined EID and excitation induced red shift of AX, consistent with the result in Fig. 2. Slight tilting of the positive peak, also seen for the $T = 36$ fs map in Fig. 3b, indicates finite inhomogeneous broadening, possibly originating from local strain and disorder.

Interestingly, at the same detection energy but higher excitation energy of $E_{ex} = 1.71$ eV, we also see strong signals. The presence of such off-diagonal peaks at ($E_{ex} = 1.71$ eV, $E_{de} = 1.64$ eV) in 2DES suggests the coupling of AX to a higher energy state. We use a notation in the bracket that the first and second symbols denote the excitation and detection energy, respectively. If we examine carefully, we indeed find that there are much weaker diagonal peaks around $E_{ex} = E_{de} = 1.7$ eV. According to literature,[54-56] we assign this peak to the excited Rydberg exciton A* of bulk $WSe_2$. Cross section spectra clearly show this diagonal peak, marked by purple arrow in Fig. 3e. The A* peak is not often seen in extensively studied TMD monolayers as its oscillator strength is much smaller compared to AX. Instead, it becomes clearer when the linewidths are narrowed by hBN encapsulation and at low temperatures.[54, 56] With increasing thickness of TMD materials, the oscillator strength of A* gradually grows, and its energy shifts closer to that of AX.[54, 55] Indeed, a low-temperature reflectance spectrum of the 60 nm thick flake confirms the presence of the A* Rydberg exciton (Supplementary Fig. 8). We do not resolve clear A* signatures in room-temperature steady-state reflectance spectra (Fig. 1b and 1d) likely due to the mixing effects of large line width and inhomogeneous broadening. The spectra are, however, indeed asymmetric and suggest the presence of A* at the higher energy side of AX (Supplementary Fig. 3). 2DES resolves such inhomogeneous broadening and overlap in linear spectrum and allows us to not only observe the much weaker A* response but also uncover the coupling between AX and A*.

With increasing waiting times, the individual distinct peaks gradually merge. For $T = 96$ fs, the overall 2DES line shape has evolved into vertical negative-positive-negative stripes centered at $E_{de}$ = AX (Fig. 3c), indicating coupling between AX and A* and other higher lying states.[49, 57]

The dynamics of (A, A), (A*, A*) and (A*, A) peaks (Fig. 4f) all oscillate with a ~50-fs period (~80 meV, see FT results in Supplementary Fig. 9), agreeing generally well with the energy difference between AX and A* extracted from 2DES. These results suggest the coherent coupling between AX and A*, i.e., the ground (n = 1) and excited (n = 2) states of the same exciton, respectively. The observation of quantum beating between these states has important implication that the Rydberg excitons can be a source for coherent control.[58]

We then employed 2DES to disentangle the spectrally congested response of the self-hybridized EPs (Fig. 4). The energy of LP, AX, and UP are marked by the black, red and blue dashed lines in the maps, respectively. At $T = 0$ fs, we observe diagonal peaks at (LP, LP) and (UP, UP) and off-diagonal peaks between LP and UP (Fig. 4a). The diagonal (LP, LP) peak is dispersive, i.e., positive-negative, as more clearly seen from the cross section in Fig. 4e (black line). Instead, (UP, UP) peak is also dispersive but with a flipped sign (Fig. 4g, black line). This line shape is a signature of Rabi contraction (Supplementary Fig. 5c) due to reduction of effective exciton-cavity coupling from PB. The PB contribution is dominant at early waiting times, consistent with the TR results in Fig. 2.

Additionally, we can also identify diagonal (AX, AX) and off-diagonal (LP, AX), (AX, LP), (UP, AX), and (AX, UP) peaks. That is, the 2DES map is composed of a set of 9 peaks, which is more obvious at $T = 12$ fs and 36 fs, even though some of them partially overlap with each other.

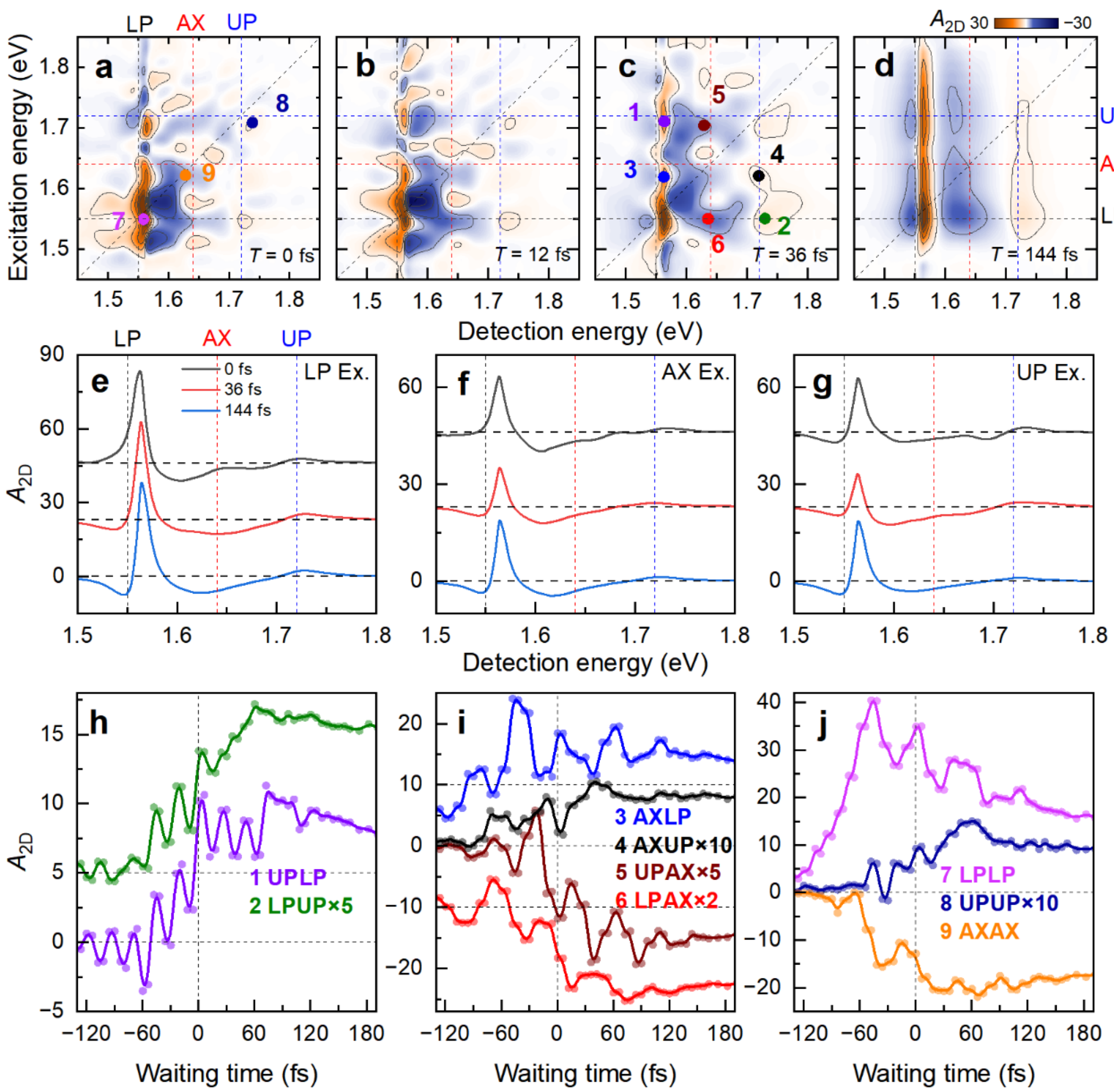

**Fig. 4 Coherent dynamics of self-hybridized exciton-polaritons.**

**a-d** 2DES maps of the 190 nm thick $WSe_2$ flake for selected waiting times of 0 fs (**a**), 12 fs (**b**), 36 fs (c), and 144 fs (**d**), revealing a significant evolution of the line shape within tens of femtoseconds. UP, AX and LP energies are marked by black, red, and blue dashed lines, respectively. **e-g** Cross section spectra taken at excitation energies of LP (**e**), AX (**f**) and UP (**g**) at waiting times $T$ = 0, 36, and 144 fs, revealing a rapid change from PB dominant to EID dominant line shapes. **h-j** Waiting-time dynamics of the selected peaks 1-9 indicated by circles in **a** and **c**, corresponding to ($E_{ex}$, $E_{de}$) of (UP, LP), (LP, UP), (AX, LP), (AX, UP), (UP, AX), (LP, AX), (LP, LP), (UP, UP) and (AX, AX) peaks.

The 2DES line shape undergoes a rapid evolution in the first 30 fs.

Specifically, the (LP, LP) peak flips from positive-negative to negative-positive-negative line shape at $T$ = 12 fs (Fig. 4b). After ~ 48 fs (Supplementary Fig. 10), the 2DES signal essentially shows positive vertical stripes at detection energies of LP and UP, and negative vertical stripes in-between LP and UP and below LP, see Fig. 4d for $T$ = 144 fs. Such a negative-positive-negative line shape is characteristic of EID. The cross sections taken at the LP, AX, and UP excitation energies are now almost the same (Fig. 4e-g, blue lines).

The 2DES results of the 60 nm thick $WSe_2$ (Fig. 2) suggest a non-negligible EIS (specifically redshift) contribution. Accordingly, we expect that self-hybridized EPs also possess a similar effect induced by the excitonic component. Indeed, comparing the red and blue lines in Fig. 4e, one observes that the positive peak of the red line is symmetric, whereas that of the blue line is asymmetric. The negative wing at the low energy side becomes stronger. This suggests the presence of EIS (redshift), becoming dominate at long waiting time.

We now examine the dynamics of the 2DES peaks (Fig. 4h-j). The cross peaks (UP, LP) and (LP, UP) show fast oscillation with a period of about 24 fs (Fig. 4h), matching the expected Rabi period of 22 fs. Strikingly, we also see strong oscillations for the (AX, LP), (AX, UP), (UP, AX) and (LP, AX) peaks in Fig. 3i, giving a period of ~ 50 fs, matching well with the energy difference of ~ 80 meV between AX and LP/UP. This 50 fs-oscillation is also seen for the (AX, AX) peak dynamics in Fig. 4j. Furthermore, the diagonal (LP, LP) and (UP, UP) peaks also show oscillations. The observation of these oscillations of coupling peaks confirms that these three states are coherently coupled.

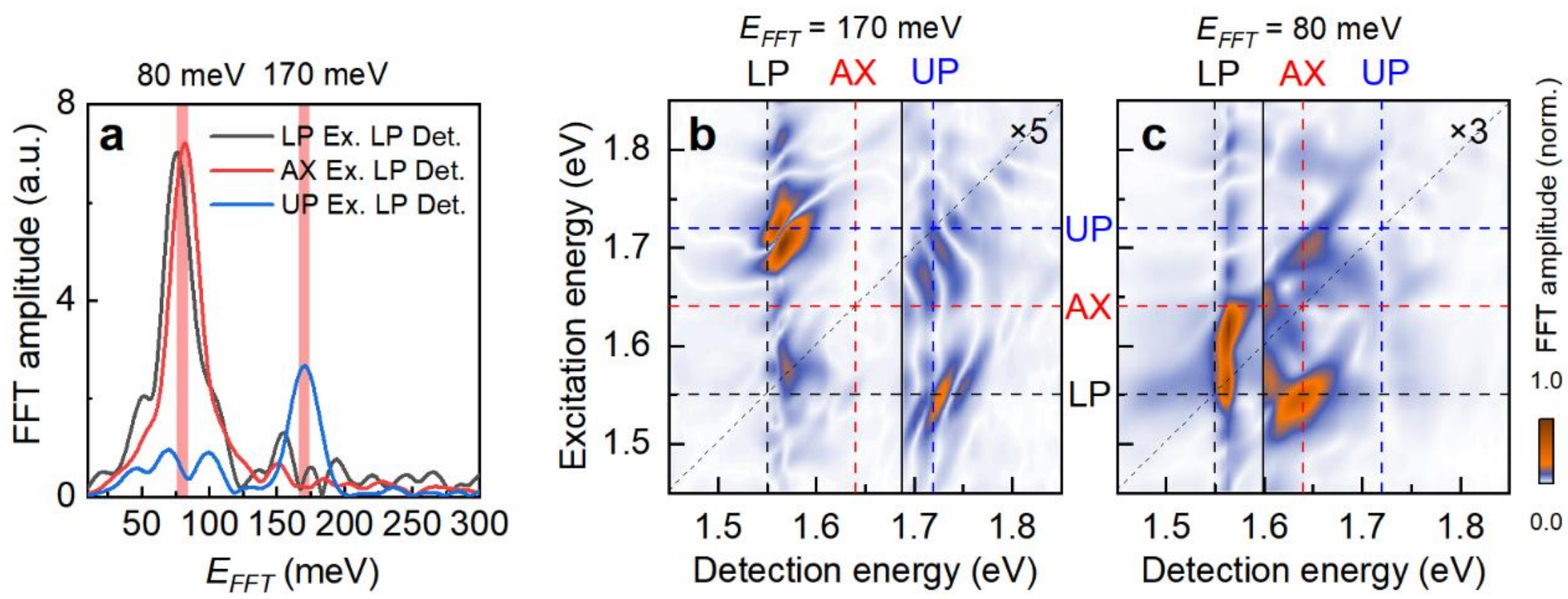

**Fig. 5 Fourier transformation along the waiting time *T*.**

**a** FFT spectra obtained by fast Fourier transformation along the $T$ axis. The spectra are taken at $E_{\mathrm{de}}$ = LP (1.56 eV) and different $E_{\mathrm{ex}}$ of LP, AX, and UP energy. The frequencies of the two main FFT peaks are determined to be $E_{\mathrm{FFT}}$ = 80 meV and $E_{\mathrm{FFT}}$ = 170 meV. **b-c** 2D beating maps obtained by performing FFT on the full-dimensional 2DES, at two different $E_{\mathrm{FFT}}$ energies of 170 ± 2 meV (**b**) and 80 ± 2 meV (**c**). The data for $E_{\mathrm{de}} > 1.68$ eV (**b**) and $E_{\mathrm{de}} > 1.6$ eV (**c**) are magnified by 5 and 3 times, respectively, for better visualization.

To further understand the frequency and origin of these oscillations, a fast Fourier transform (FFT) analysis of the dynamics along $T$ is performed. FFT spectra at $E_{de}$ = LP and $E_{ex}$ = LP, AX, and UP are presented in Fig. 5a (complete data can be found at Supplementary Fig. 11). Two main peaks at $E_{FFT}$ = 170 meV and $E_{FFT}$ = 80 meV are seen. A 2D beating map at $E_{FFT}$ = 170 meV (Fig. 5b) shows pronounced peaks at the two diagonal and two cross peaks between LP and UP. This confirms that this fast oscillation is the expected Rabi oscillation arising from the coherent population transfer between AX and the cavity mode. The lower energy oscillation at $E_{FFT}$ = 80 meV exhibits strong amplitudes at cross peaks between AX and UP/LP, thus reflecting their coherent coupling feature (Fig. 5c).

We note that we do not clearly resolve the diagonal and cross peaks of A* in the 2DES spectra of the 190 nm thick sample. This is reasonable since the coupling between A* and the cavity is much weaker than that of AX due to its substantially lower oscillator strength (Fig. 3 and Supplementary Fig. 3) and its resonance is far detuned from the cavity (~100 meV). Therefore, the 2DES results of the 190 nm thick sample mainly reflect the coupling of FP cavity mode and AX, with much weaker contributions of A*.

Furthermore, we observe that all coherent oscillations decay almost completely by $T = 120$ fs, marking the transition from predominantly coherent to incoherent dynamics. Together with the rapid line shape evolution, this observation evidences the transition from coherent polaritons to incoherent polariton excitations.[10]

**Three coupled oscillator model simulation and interpretation of 2DES**

Such a coupling between polaritons and "exciton" has previously been seen in molecular J-aggregates coupled to plasmonic grating cavities.[44] This is attributed to the spatially inhomogeneous plasmonic field that leads to strong and weak coupling with molecular excitons. By calculating the electromagnetic field distribution within the FP cavity modes, we find that the cavity electromagnetic field $E(d)$ at different depth $d$ is also inhomogeneous (Supplementary Fig. 12), agreeing with the expected 2$^{nd}$ order cavity mode profile with two nodes. Thus, excitons with a transition dipole moment $\mu_X$ in different depths of the flake will have different coupling strength with the cavity because $g(d) = \mu_X E(d)$.[59, 60]

To simply the model, we assume that there are two sets of excitons that couple strongly ($|X_S\rangle$) and weakly ($|X_W\rangle$) to the same cavity mode $|C\rangle$ (schematically shown in Fig. 6a).[44] Note that each subset of excitons essentially reflects many excitons (*N*) in the flake that are collectively coupled to the same cavity mode. We then reach a minimum three-coupled oscillator model (3-COM), including $|C\rangle$, $|X_S\rangle$, and $|X_W\rangle$, which are the one-quantum (1Q) states of the system (Supplementary Fig. 13). In addition to the regular LP and UP, this results in an additional hybridized state residing at the exciton energy, which we refer to as a "dark" state $|D\rangle$ (Fig. 6b), whose wavefunction is $|D\rangle = -C_{X_S,D}|X_S\rangle + C_{X_W,D}|X_W\rangle$, where $C_{X_S,D}$ and $C_{X_W,D}$ are the corresponding coefficients. That is, $|D\rangle$ is

purely excitonic and does not contain a cavity component, like the ($N$-1) dark states in the Tavis-Cummings model.[61] $|D\rangle$ will thus reside at the AX energy. We note that a similar 3-COM has been used to describe monolayer TMD excitons coupled to a plasmonic grating nanocavity[10] and self-hybridized EPs enabled by TMD metasurfaces.[9] Here, the strong spatial near-field localization in these samples allows the coupling to both momentum bright and dark excitons.

We then construct an effective Hamiltonian based on this 3-COM and calculate 2DES spectra by analyzing the time evolution of the density matrix by solving the Lindblad master equation (Methods). For describing the nonlinear third-order 2DES, it is essential to include the 2Q states, i.e., doubly excited cavity $|2C\rangle$, two-exciton states $|2X_S\rangle$ and $|2X_W\rangle$, and three mixed states $|C, X_S\rangle$, $|C, X_W\rangle$ and $|X_S, X_W\rangle$ (Supplementary Fig. 13). This results in 6 hybridized states in 2Q (Fig. 6b). Analysis of GSB/SE and ESA involving the transitions between the ground state $|0\rangle$, 1Q and 2Q states will thus give us 2DES line shapes. Importantly, when considering all the three 1Q states as harmonic oscillators, no optical nonlinearity of the coupled system is expected, i.e., no signal in TR and 2DES. We assume that the FP cavity is a harmonic oscillator. For the TMD excitons, MBI lead to a strong optical nonlinearity. MBI are implemented in our model by modifying the properties of 2Q states, following the approaches in ref.[10, 51, 62]

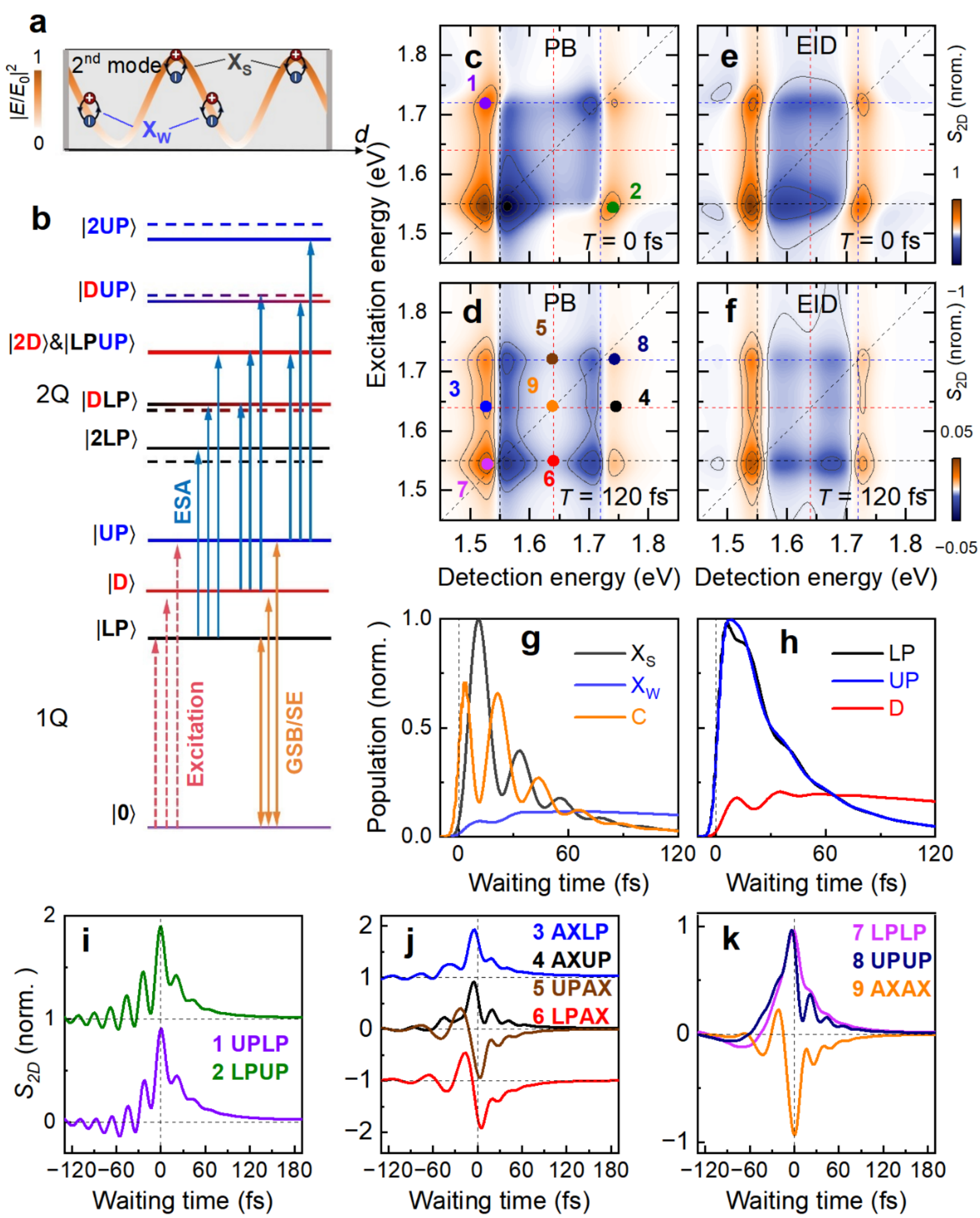

**Fig. 6 Simulations of 2DES maps with different MBIs effects.**

**a** Schematic diagram of coupling between the FP cavity mode with a spatially inhomogeneous electromagnetic field and excitons, which include strongly ($|X_S\rangle$) and weakly ($|X_W\rangle$) coupled excitons. *d* represents for the thickness. **b** Eigenstate structure of the 3-COM Hamiltonian including PB effect in the coherent dynamics regime. The coupling of $|X_S\rangle$ and $|X_W\rangle$ to the FP cavity mode $|C\rangle$ gives rise to two polariton states, $|LP\rangle$ and $|UP\rangle$, and a dark state $|D\rangle$. The pump pulse excites all three 1Q species and the probe pulse gives positive GSB/SE signals (orange arrows) in 2DES. The blue arrows represent the transitions to the 2Q manifold with six different states, resulting in negative ESA signals in 2DES. **c-f** Simulated 2DES maps at waiting time $T$ = 0 fs and 120 fs with PB (**c**-**d**) and EID (**e**-**f**) effects. **g-h** Population dynamics in the exciton-cavity basis **(g)** and polariton basis **(h)**. **i-k** Waiting time dynamics of the nine diagonal and cross peaks, extracted from the positions marked by 1-9 in **c** and **d**.

We first consider the effect of PB, i.e., a reduction in transition dipole moment of the $|X\rangle \rightarrow |2X\rangle$ transition relative to $|0\rangle \rightarrow |X\rangle$ transition. An optical excitation of excitons reduces the number of excitons that are coupled to the cavity, thus reducing the normal mode splitting. This effect is known as Rabi contraction and has been observed in exciton-polaritons and vibration-polaritons.[10, 37, 52, 53] In our model, this results in a blue shift of $|2LP\rangle$ and $|D, LP\rangle$ and red shift of $|2UP\rangle$ and $|D, UP\rangle$ (Fig. 6b). This energy shift induces an optical nonlinearity of polaritons and has consequences in determining the line shape of the peaks. Specifically, the diagonal (LP, LP) peak has contribution from the GSB/SE of $|0\rangle \rightarrow |LP\rangle$ and ESA of $|LP\rangle \rightarrow |2LP\rangle$ transitions. Blue shift of $|2LP\rangle$ energy thus leads to dispersive positive-negative line shape for (LP, LP). In contrast, red shift of $|2UP\rangle$ leads to a dispersive negative-positive line shape for (UP, UP). This is seen in the simulated 2DES map at $T$ = 0 fs (Fig. 6c) that matches well with the diagonal peak in experiment. At $T$ = 0 fs, the cross peaks of (LP, UP) and (UP, LP) are mostly absorptive, which turns to dispersive at larger $T$ (Supplementary Fig. 15).

We then turn to the effect of EID, which is included in our model by increasing the pure dephasing rate of the $|X\rangle \rightarrow |2X\rangle$ transition relative to that of $|0\rangle \rightarrow |X\rangle$. As seen in the simulated 2DES map (Fig. 6e and 6f), this gives negative-positive-negative line shape for all the diagonal and cross peaks of LP and UP, which generally reproduce the data for long waiting times.

We note that, however, the line shape for $T > 36$ fs cannot be fully explained by PB and EID effects. As discussed above, we see a relatively strong negative signal at the detection energy slightly lower than LP for all excitation energies (Fig. 4e-g). In the simulation, however, the negative signal at this detection energy is relatively weak even for EID, and PB does not give this negative signal at all. Such a stronger negative wing at the low energy side of a resonance is seen in the 60 nm thick sample and is due to EIS (specifically redshift). We thus also consider EIS in our model and the simulated 2DES maps are shown in Supplementary Fig. 15, clearly showing the dispersive negative-positive signal for all the diagonal and cross peaks. Thus, we attribute the line shape at late waiting times to the combined effect of EIS and EID. The simulation results at longer waiting times (Fig. 6d,6f, more data in Supplementary Fig. 15) reproduce generally well the vertical stripes and features of diagonal peak of (AX, AX) and the cross peaks between AX and UP/LP.

After analyzing the line shape, we now examine the dynamics. The simulation indicates strong coherent population transfer between $|X_S\rangle$ and $|C\rangle$ and a relatively weak oscillation of the population of $|X_W\rangle$ (Fig. 6g). In the polariton basis (Fig. 6h), we observe reversible energy exchange between dark states and polaritons. After about 100 fs, the population resides mainly in the dark states, indicating the dephasing of polaritons to the dark states.

Fig. 6g-i shows the dynamics of different diagonal and cross peaks, taken at the energy positions marked by numbers 1-9 in Fig. 6c and 6d. (UP, LP) and the (LP, UP) peaks show oscillations of 22-fs period. The oscillation period of the (AX, LP/UP), (LP/UP, AX) and (AX, AX) signal increases to about 44 fs due to a lower energy difference between AX and LP. The diagonal (LP, LP) peak shows weaker oscillation. All these results reproduce well the experimental findings. Similar dynamics are also seen for EID and EIS effects (Supplementary Fig. 16).

The simulation shows that the coherent oscillations fade away at ~100 fs, agreeing well with the experiment. Importantly, this loss of coherent oscillation is accompanied by a clear evolution of the line shape (Fig. 4a-d and Supplementary Fig. 10). Immediately after optical excitation, the line shape is dominated by Rabi contraction, a hallmark of polaritons. Since coherent polaritons carry a photonic component (or less excitonic component compared to pure excitons), the interactions between excited-state quasiparticles are substantially weakened. Rabi contraction induced by PB is independent of these excited-state quasiparticle interactions. Rabi contraction thus dominates the nonlinearity of coherent polaritons around $T = 0$ fs. With increasing waiting time, coherent polaritons gradually lose their photonic character and decay to collective dark states. Since these dark states carry no photonic component, quasiparticle interactions such as exciton-exciton scattering (leading to EID) and Coulomb interactions (leading EIS) progressively grow in strength. Thus, EID and EIS nonlinearities become dominant. Therefore, as the waiting time increases, the growing population of the dark states drastically enhances MBI. Combing the dynamics and line shape, our results capture the complete transition from coherent to incoherent polariton excitations.[9, 10]

Our results indicate that coherent polariton dynamics can be achieved at room temperature in readily available semiconductor flakes supporting self-hybridized EPs, significantly expanding the scope of candidate materials for a variety of applications like ultrafast switching.

## Conclusion

In summary, we demonstrate ultrafast coherent Rabi oscillation in an exfoliated thin flake of $WSe_2$ with sub-wavelength thickness, enabled by strong coupling of excitons and the FP cavity mode. The 2DES measurements show cross peaks not only between LP and UP but also between these polariton branches and excitons, all showing coherent amplitude oscillations. These results can be rationalized by excitons that either experience strong or weak coupling to the FP cavity mode, depending on their position within the cavity because of the inhomogeneous cavity-field distribution. The 2DES line shape undergoes a dramatic evolution within the first 100 fs, suggesting a change in optical nonlinearity from Rabi contraction of polaritons to MBIs of the excitonic system, corresponding to transition from coherent to incoherent excitations of polaritons. We further observe clear signatures of quantum beating between the Rydberg exciton (A*) and AX. Our results provide insight into the ultrafast coherent dynamics of self-hybridized EPs and highlight the potential of readily accessible, subwavelength-thick $WSe_2$ flakes for coherent control in external-cavity-free polaritonics.

# Methods

## Sample preparation

All $WSe_2$ flakes were mechanically exfoliated on PDMS and then transferred on to an Au/$SiO_2$/Si substrate by a two-dimensional material transfer platform (Metatest E1-T). The Au layer was deposited on $SiO_2$/Si substrate by electron beam evaporation. The thickness of the $WSe_2$ flakes is characterized using atomic force microscopy (AFM, Dimension Edge, Bruker), and the results are displayed in Supplementary Fig. 1.

## Transient reflection and 2DES measurements

Ultrafast spectroscopic measurements were conducted in a home-built setup in a pump-probe configuration (Supplementary Fig. 6). All experiments were performed in the linear regime (Supplementary Fig. 7). A Yb-YAG laser source (CARBIDE CB-29W, Light Conversion) delivers 1030 nm laser pulse with 180 fs pulse duration at a repetition of 100 kHz, which pumps a home-built noncollinear optical parametric amplifier (NOPA, detail configurations in Supplementary Fig. 6a). The NOPA outputs broadband 14 fs laser pulse (640-880 nm), which is separated and used as both the pump and probe pulses. The laser spectrum is shown in Supplementary Fig. 6d. The pump and probe pulses then pass through two mechanical chopper systems (MC2000B-EC, Thorlabs) separately with customized blades to operate at 25 kHz (pump) and 50 kHz (probe) chopping rate for double modulation lock-in detection to suppress pump scattering.[63]

For 2DES experiment (Supplementary Fig. 6b), by passing the pump through an interferometer based on birefringent wedges (GEMINI 2D, NIREOS), two collinear, phase-locked replicas of the pump pulse[64] with variable time delay τ (coherence time) are generated. The noncollinear probe pulse arrives at the sample after a waiting time $T$ with respect to the second pump pulse, which is controlled by a motorized delay line (MFA-CC, Newport). The pump pulses pair and probe pulse were focused on the sample at θ = 5.7° by a concave mirror with diameters ~20 μm. The reflected probe from sample was recorded as a function of coherence and waiting time. The spectra were detected by a spectrometer that is equipped with a monochromator (SP-2150, Princeton Instruments) and a high acquisition rate fast line camera (Ultraspec-VISDL/SL-128-100K, TrSPEC), which gives the detection energy ($E_{de}$). The differential reflection spectra were obtained by:

$$\frac{\Delta R}{R}(\tau, T, E_{de}) = \frac{R_{on}(\tau,T,E_{de}) - R_{off}(E_{de})}{R_{off}(E_{de})} \qquad (1)$$

in which $R_{on}$ and $R_{off}$ represent the reflection spectrum in the presence and absence of the pump pulse, respectively. Transient differential reflection spectra were obtained by setting $\tau = 0$.

To acquire the excitation energy ($E_{ex}$) axis, the coherence time $\tau$ is scanned from −300 to 500 fs for each waiting time *T*. By taking the real part of the Fourier transform of the differential reflection spectra along $\tau$, absorptive 2DES maps were then obtained:[44]

$$A_{2D}(E_{ex},T,E_{de}) = \mathcal{R}\left(\int_0^\infty \frac{\Delta R}{R}(\tau,T,E_{de})e^{iE_{ex}\tau/\hbar}\,d\tau\right) \quad (2)$$

**Dynamics fitting and frequency-domain analysis**

The 2DES peak dynamics consist of a slowly varying non-oscillatory background $y_{base}(t)$ superimposed on coherent oscillatory components $y_{osc}(t)$. To separate the background from the oscillatory components, we employ a Savitzky-Golay (SG) filtering approach.[65] SG smoothing with a window length of 19 fs and fit to a third-order polynomial are applied to extract the slow non-oscillatory baseline $y_{base}(t)$ from the raw experimental data.[66] The coherent oscillatory signal $y_{osc}(t)$ is then obtained by subtracting this fitted slow background from the raw peak dynamic traces. For frequency-domain analysis, the extracted oscillatory signal $y_{osc}(t)$ is multiplied by a Gaussian window function to suppress edge artifacts, followed by detrending and zero-padding. Fast Fourier transform (FFT) is performed to resolve the oscillation frequency $E_{FFT}$.

## TMM simulation

The FP cavity mode with finite thickness $WSe_2$ is calculated by TMM simulation.[67] To simplify the calculation, the gold substrate is assumed to be infinitely thick. The characteristic matrix $M_j$ for the *j*-th layer is

expressed as:

$$\mathrm{M}_j = \begin{bmatrix} \cos(\varphi_j) & -i\frac{1}{S_i}\sin(\varphi_j) \\ -iS_j\sin(\varphi_j) & \cos(\varphi_j) \end{bmatrix} \tag{3}$$

Here,

$$S_j = \frac{\sqrt{\frac{\varepsilon_0}{\mu_0}}n_j}{cos(\theta_j)} \tag{4}$$

Where $\varepsilon_0$ and $\mu_0$ are the vacuum permittivity and vacuum permeability, respectively, $n_j$ is the complex refractive index of the $j$-th medium, and $\theta_j$ is the angle between the wave vector and the normal inside the $j$-th layer. It should be noted that the subscript "$j$" here takes the values of 0, 1, 2, and "s", which denote the air, the dielectric layer, the gold film layer, and the Si substrate, respectively. For simplicity, $\theta_j$ is formally obtained from $n_0\sin\theta = n_j\sin\theta_j$, with $\theta$ the angle of incidence and $n_0$ the refractive index of air. $\varphi_j = \frac{2\pi}{\lambda}n_j d_j cos(\theta_j)$, is the phase shift angle experimented by the wave after crossing the layer $j$, with $d_j$ the layer thickness. The total characteristic matrix of the system is $\mathrm{M} = \mathrm{M}_1 * \mathrm{M}_2$. The reflectance spectrum of the sample is calculated using the formula:

$$\mathrm{R} = \left|\frac{\mathrm{S_0(M(1,1)+S_sM(1,2)-(M(2,1)+S_sM(2,2))}}{\mathrm{S_0(M(1,1)+S_sM(1,2)+(M(2,1)+S_sM(2,2))}}\right|^2 \tag{5}$$

**Density matrix simulation**

To simulate the linear and nonlinear spectra, we numerically solve the Lindblad master equation[68, 69] in a non-perturbative approach and obtain the time evolution of the density matrix $\hat{\rho}(t)$

$$\dot{\hat{\rho}} = -\frac{i}{\hbar}\left[\hat{H}, \hat{\rho}\right] + \frac{1}{2}\sum_k (2\hat{L}_k \hat{\rho} L_k^\dagger - L_k^\dagger \hat{L}_k \hat{\rho} - \hat{\rho} L_k^\dagger \hat{L}_k). \qquad (6)$$

The full Hamiltonian $\hat{H} = \hat{H}_S + \hat{H}_{int}$ contains the system Hamiltonian and its light-matter interaction part $\hat{H}_{int} = -\hat{\mu}E(t)$, which accounts for the optical excitation of the system by an external light field $E(t)$ in the dipole approximation. Here $\hat{\mu}$ is the total transition dipole moment of the system ($\hat{\mu} = \hat{\mu}_C + \hat{\mu}_{XS} + \hat{\mu}_{XW}$), and $E(t) = E_{\mathrm{pu}}(t) + E_{\mathrm{pr}}(t)$ is the electric field of the pulse impinging on the sample, comprising both the pump $E_{\mathrm{pu}}(t)$ and probe $E_{\mathrm{pr}}(t)$ pulses,

$$E(t) = \sum_n E_{0,\mathrm{n}} \exp\left(-2\ln 2\left(\frac{t-t_n'}{\Delta \mathrm{t}}\right)^2\right)\cos(\omega_L(t - t_n') + \phi_n) \quad (7)$$

where *n* runs over three laser pulses (pump 1, pump 2, and probe) with amplitude $E_{0,\mathrm{n}}$, pulse duration $\Delta \mathrm{t} = 6$ fs, laser frequency $\omega_\mathrm{L} = 1.639\ \mathrm{eV}/\hbar$ with $\hbar$ being the reduced Planck constant, phase $\phi_\mathrm{n}$ for phase cycling and a time shift $\mathrm{t}_\mathrm{n}'$. The delay between the two pump pulses is the coherence time $\tau = \mathrm{t}_{\mathrm{pu2}}' - \mathrm{t}_{\mathrm{pu1}}'$ and the delay between the second pump pulse and the probe pulse is the waiting time $\mathrm{T} = \mathrm{t}_{\mathrm{pr}}' - \mathrm{t}_{\mathrm{pu2}}'$. For calculating pump-probe spectra, we set $\tau = 0$ in the simulations.

After obtaining the time evolution of $\hat{\rho}(t)$, the polarization is calculated as the expectation value of the transition dipole moment via $\mathrm{P}(t) = \mathrm{Tr}(\hat{\mu}\hat{\rho}(t))$, which gives us the optical absorption spectra of the system. We calculate the total polarization $\mathrm{P}_T(\tau, T, t)$ and linear polarization $P_L(t)$ in the presence and absence of the pump pulse, respectively. Specifically, the total polarization is calculated via a phase-cycling method to get the emission in the probe direction as detected in experiment.[70, 71] The linear polarization is subtracted from the total polarization to get the pure nonlinear polarization $P_{NL}(\tau, T, t) = P_T(\tau, T, t) - P_L(t)$ . The nonlinear susceptibility is obtained via Fourier transform after referencing to the probe pulse field,

$$\chi_{NL}(\tau, T, E_{\mathrm{de}}) = \frac{1}{\varepsilon_0} \frac{\mathcal{F}(P_{NL}(\tau, T, t))}{\mathcal{F}\left(E_{\mathrm{pr}}(t)\right)} \quad (8)$$

where $\varepsilon_0$ denotes the vacuum dielectric constant and $E_{\mathrm{de}}$ is the probe/detection energy. The nonlinear spectra are then taken as the imaginary part,

$$S_{NL}(\tau, T, E_{\mathrm{de}}) = \Im\left(\chi_{NL}(\tau, T, E_{\mathrm{de}})\right). \quad (9)$$

Finally, 2DES maps for a fixed waiting time $T$ are obtained by taking the real part of the Fourier transform along the coherence time $\tau$

$$S_{NL}(E_{\mathrm{ex}}, T, E_{\mathrm{de}}) = \Re\left(\int_{-\infty}^{\infty} \theta(\tau) S_{NL}(\tau, T, E_{\mathrm{de}}) e^{iE_{\mathrm{ex}}\tau/\hbar} d\tau\right) \quad (10)$$

where $\theta(\tau)$ denotes the Heaviside step function. This yields the excitation energy axis $E_{\mathrm{ex}}$ and thus form 2DES maps as a function of $E_{\mathrm{ex}}$, $E_{\mathrm{de}}$ and $T$.

The system Hamiltonian based on the 3-COM is

$$\hat{H}_S = E_C(\theta)\hat{b}_C^\dagger \hat{b}_C + \sum_{X=X_S,X_W} E_X \hat{b}_X^\dagger \hat{b}_X + \sum_{X=X_S,X_W} V_X\left(\hat{b}_C^\dagger \hat{b}_X + \hat{b}_X^\dagger \hat{b}_C\right) \quad (11)$$

where FP cavity is considered as a harmonic oscillator with creation and annihilation operations $\hat{b}_C^\dagger$ and $\hat{b}_C$, respectively. Angle-dependent FP cavity energy $E_C(\theta)$ is deduced from TMM simulations. As discussed in the main text, in this phenomenological model, we include two types of excitons ( $|X_S\rangle$ and $|X_W\rangle$ ) with exciton creation and annihilation operations $\hat{b}_X^\dagger$ and $\hat{b}_X$, respectively. Note that these are also bosonic harmonic oscillator operators. The exciton energy $E_X = 1.64\ eV$ is the same for these two types of excitons. The exciton-cavity coupling is implemented in rotating wave approximation and the coupling strength for the two types of excitons is different, i.e., $V_W < V_S$.

The interaction of exciton and cavity with external laser field is accounted for by using the transition dipole moment operators as $\hat{\mu}_{X/C} = \mu_{X/C}(\hat{b}_{X/C}^\dagger + \hat{b}_{X/C})$. We introduce exciton dephasing via the Lindblad operator $\hat{L}_{dep,X} = \sqrt{2\gamma_X}\hat{b}_X^\dagger\hat{b}_X$. The radiative damping of FP cavity is included as $\hat{L}_C = \sqrt{\kappa_C}\hat{b}_C$. This gives cavity lifetime $T_{1,C} = 1/\kappa_C$. Note that we do not include exciton depopulation as the relevant time scale is much larger than our experimental time delay window.

## Author Contributions

J.H.Z. supervised the project. N.X., J.Y. and K.L. conceived the idea and designed the research. K.L., J.Y. prepared the samples and measured the thickness. N.X. J. Y. and J. S. constructed the NOPA and 2DES system, and performed the ultrafast experiments, and analyzed the data under the supervision of J.H.Z., K.L. set up the angle-resolved spectroscopy and completed the steady-state spectra acquisition. The TMM simulations were performed by K.L. and N.X. The density matrix simulations were performed by J.H.Z. with the help of D.T. and A.C. The manuscript was written by N.X., J.Y., K.L., W.S. and J.H.Z. All authors contributed to discussions or gave feedback to the final version.

## Acknowledgements

We thank Christoph Lienau for helpful discussion. This research has been financially supported by National Key R&D Program of China (2024YFA1210803), National Natural Science Foundation of China (22272072, W2412113), Guangdong Basic and Applied Basic Research Foundation (2023ZT10X010, 2025A1515012266), Shenzhen Key Laboratory of Phononics and Intelligent Thermal Materials (SYSRD20250529114001002), and High level of special funds (G03050K002). A.C. is thankful to Iran National Science Foundation (INSF) for the financial support provided under the grant number 4030055.